\documentclass[a4paper,12pt,doublespacing]{article}
\usepackage{fullpage,setspace}
\usepackage[hmargin=1in,vmargin=1.5in]{geometry}
\usepackage{amsfonts,amsmath,amsthm}
\usepackage[latin1]{inputenc}
\usepackage{textcomp}
\usepackage{version,amsfonts,amsmath,boxedminipage,alltt}
\usepackage{graphics}
\usepackage{graphicx}
\usepackage{multicol}
\usepackage{ctable}
\usepackage{multirow}

\theoremstyle{remark}

\let\epsilon=\varepsilon

\let\phi=\varphi

\newcommand{\E}{\field{E}}

\newcommand{\field}[1]{\mathbb{#1}}

\def\der^#1_#2{\frac{\partial^{#1}}{\partial {#2}^{#1}}}

\usepackage[eurosym,right]{eurofont}

\title{Large tick assets: implicit spread and optimal tick size}
\author{Khalil Dayri\\ Antares Technologies\\ khalil.al-dayri@polytechnique.edu\and
Mathieu Rosenbaum\\ Laboratory of Probability and Random Models,\\
University Pierre and Marie Curie (Paris 6)\\
mathieu.rosenbaum@upmc.fr}

\date{\today}
\begin{document}
\maketitle

\begin{abstract}
In this work, we provide a framework linking microstructural properties of an asset to the tick value of the exchange. In particular, we bring to light a quantity, referred to as {\it implicit spread}, playing the role of spread for large tick assets, for which the effective spread is almost always equal to one tick. The relevance of this new parameter is shown both empirically and theoretically. This implicit spread allows us to quantify the tick sizes of large tick assets and to define a notion of {\it optimal tick size}. Moreover, our results open the possibility of forecasting the behavior of relevant market quantities after a change in the tick value and to give a way to modify it in order to reach an optimal tick size.
\end{abstract}


\section{Introduction}\label{intro}

\subsection{Tick value, tick size and spread}

On electronic markets, the market platform fixes a grid on which traders can place their prices. The grid step represents the smallest interval between two prices and is called the \textit{tick value} (measured in the currency of the asset). For a given security, it is safe to consider this grid to be evenly spaced even though the market may change it at times. In some markets, the spacing of the grid can depend on the price. For example, stocks traded on Euronext Paris have a price dependent tick scheme. Stocks priced $0$ to $9.999$ euros have a tick value of $0.001$ euro but all stocks above $10$ euros have a tick value of $0.005$ euro.

However, when it comes to actual trading, the tick value is given little consideration. What is important is the so called \textit{tick size}. A trader considers that an asset has a small tick size when he ``feels" it to be negligible, in other words, when he is not averse to price variations of the order of a single tick. In general then, the trader's perception of the tick size is qualitative and empirical, and depends on many parameters such as the tick value, the price, the usual amounts traded in the asset and even his own trading strategy. All this leads to the following well known remark : the tick value is not a good absolute measure of the perceived size of the tick. It has to be viewed relatively to other market statistics. For instance, every trader ``considers" that the ESX index futures has a much greater tick than the DAX index futures though their tick values have the same order of magnitude.

Nevertheless, the notion of ``large tick asset'' is rather well understood. For Eisler, Bouchaud and Kockelkoren \cite{ebk}: ``large tick stocks are such that the bid-ask spread is almost always equal to one tick, while small tick stocks have spreads that are typically a few ticks". We borrow this definition in this work. This type of asset lead to the following specific issues which we address in this paper:

\begin{itemize}
\item How to quantify more precisely the tick sizes of large tick assets ?
\item Many studies have pointed out special relationships between the spread and some market quantities. However, these studies reach a limit when discussing large tick assets since the spread is artificially bounded from below. How to extend these studies to this kind of asset ?
\item What happens to the relevant market quantities when the market designer decides to change the tick value and what is then the optimal tick value ?
\end{itemize}

This last question is a crucial issue facing by market designers and regulators today, see for example \cite{sec}. This is shown by the numerous changes and come back recently operated on the tick values on various exchanges. In particular, the tick value is one of the main tools the exchanges have at their disposal to attract/prevent high frequency trading. To our knowledge, this question has been surprisingly quite ignored in the quantitative financial economics literature. We believe our approach is a first quantitative step towards solving this important and intricate problem.

In this paper, in order to address the questions related to the tick value, we present a framework that allows us to link some microstructural features of the asset together. In the literature, such attempts have been considered many times and in the following two paragraphs we recall two approaches leading to important relationships between the spread and other market quantities in the case of small tick assets. However, these works focusing particularly on the spread, they are not relevant when dealing with large tick assets since in that case, the spread is collapsed to the minimum and is equal to one tick. We draw inspiration from these theories and investigate the existence of a variable that can be used in lieu of the spread in the case of large tick assets.

\subsection{The Madhavan \emph{et al.} spread theory for small tick assets}
The way the spread settles down in the market is widely studied in the microstructure literature, see for example \cite{biais,glo1,glo2,has,hol,hua,hua2,mad,ohara,sto}. In particular, several theoretical models
have been built in order to understand the determinants of the spread, see \cite{bou,dan,fou,luc,ros,smi}. Here we give a here a brief, simplified, overview of Madhavan, Richardson, Roomans seminal paper \cite{mrr} about the link between spread and volatility. In \cite{mrr}, the authors assume the existence of a {\it true} or {\it efficient price} for the asset with ex post value $p_i$ after the i$^{\text{th}}$ trade and that all transactions have the same volume. Then they consider the following dynamic for the efficient price:

\begin{equation*}
p_{i+1}-p_i=\xi_i+\theta\varepsilon_i,
\end{equation*}
with $\xi_i$ an iid centered shock component (new information,\ldots) with variance $v^2$, $\varepsilon_i$ the sign of the i$^{\text{th}}$ trade and $\theta$ an impact parameter. Note that, in order to simplify the presentation, we assume here that the $\varepsilon_i$ are independent (in \cite{mrr}, the authors allow for short term dependence in the $\varepsilon_i$).

The idea in \cite{mrr} is then to consider that market makers cannot guess the surprise of the next trade. So, they post (pre trade) bid and ask prices $a_i$ and $b_i$ given by
$$a_i=p_i+\theta+\phi,~~b_i=p_i-\theta-\phi,$$ with $\phi$ an extra compensation claimed by market makers, covering processing costs and the shock component risk. The above rule ensures no ex post regret for the market makers: If $\phi=0$, the traded price is on average the right one. In particular, the ex post average cost of a market order with respect to the efficient price $a_i-p_{i+1}$ or $p_{i+1}-b_i$ is equal to $0$.

The Madhavan {\it et al.} model allows to compute several relevant quantities. In this approach, we obtain that\footnote{We use the symbol $\sim$ for approximation.}
\begin{itemize}
\item The spread $S$ is given by $S=a-b=2(\theta+\phi)$.
\item Neglecting the contribution of the news component, see for example \cite{wbkpv} for details, the variance per trade of the efficient price $\sigma_{tr}^2$ satisfies
$$\sigma_{tr}^2=\E[(p_{i+1}-p_i)^2]=\theta^2+v^2\sim \theta^2.$$
\item Therefore:
$$S\sim 2\sigma_{tr}+2\phi.$$
\end{itemize}

This last relation, which gives a very precise link between the spread and the volatility per trade, will be one of the cornerstones of our study.

\subsection{The Wyart \emph{et al.} approach}

We recall now the Wyart {\it et al.} approach, see \cite{wbkpv}, which is another way to derive the proportionality between the spread and the volatility per trade. Here again, the idea is to use the dichotomy between market makers and market takers. Market makers are patient traders who prefer to send limit orders and wait to be executed, thus avoiding to cross the spread but taking on volatility risk.
Market takers are impatient traders who prefer to send market orders and get immediate execution, thus avoiding volatility risk but crossing the spread in the process.
Wyart {\it et al.} consider a generic market making strategy on an asset and show that its average profit and loss per trade per unit of volume can be well approximated by the formula
$$\frac{S}{2}-\frac{c}{2}\sigma_{tr},$$
where $S$ denotes the average spread and $c$ is a constant depending on the asset, but which is systematically between $1$ and $2$.

This profit and loss should correspond to the average cost of a market order. Then Wyart {\it et al.} argue that on electronic market, any agent can chose between market orders and limit orders. Consequently the market should stabilize so that both types of orders have the same average (ex post) cost, that is zero. Indeed, because of the competition between liquidity providers, the spread is the smallest admissible value such that the profit of the market makers is non negative (otherwise another market maker would come with a tighter spread). Thus, if the tick size allows for it, the spread is so that market makers do not make profit. Therefore, in this case:
$$S\sim c\sigma_{tr}.$$
Moreover, in \cite{wbkpv}, Wyart {\it et al.} show that this relationship is very well satisfied on market data.

\subsection{Aim of this work and organization of the paper}
The goal of this work is to provide a framework linking microstructural properties of the asset to the tick value of the exchange. Because the microstructure manifests itself through the statistics of the high frequency returns and durations, our approach is to find a formula connecting the tick value to these statistics. As a consequence of that, we are able to predict these statistics whenever a change in the tick value is scheduled. Furthermore, we can determine beforehand what should the value of the tick be if the market designer has a certain set of high frequency statistics he wants to achieve.

In order to reach that goal, we bring to light a quantity, referred to as {\it implicit spread}, playing the role of spread for large tick assets, for which the effective spread is almost always equal to one tick. In particular, it enables us to quantify the tick sizes of this type of asset and to define a notion of {\it optimal tick size}. The implicit spread is introduced thanks to a statistical model described in Section \ref{uz}. In order to validate the fact that our new quantity can be seen as a spread for large tick assets, we show in Section \ref{num} the striking validity of the relationship between spread and volatility per trade mentioned above on various electronically traded large tick assets, provided the spread is replaced by the implicit spread. We also explain this relationship from a theoretical point of view through a very simple equilibrium model in Section \ref{theo}. Finally, in Section \ref{change}, we show that these results enable us to forecast the behavior of relevant market quantities after a change in the tick value and to give a way to modify it in order to reach an optimal tick size.

\section{The model with uncertainty zones}\label{uz}

The implicit spread can be naturally explained in the framework of the model with uncertainty zones developed in \cite{rr1}. Note that we could introduce this notion without referring to this model. However, using it is very convenient in order to give simple intuitions.

\subsection{Statistical model}

The model with uncertainty zones is a model for transaction prices and durations (more precisely, only transactions leading to a price change are modeled). It is a {\it statistical model}, which means it has been designed in order to reproduce the stylized facts observed on the market and to be useful for practitioners. It is shown in \cite{rr1} that this model indeed reproduces (almost) all the main stylized facts of prices and durations at any frequency (from low frequency data to ultra high frequency data). In practice, this model is particularly convenient in order to estimate relevant parameters such as the volatility or the covariation at the ultra high frequency level, see \cite{rr2}, or when one wants to hedge a derivative in an intraday manner, see \cite{rr3}.

A priori, such a model is not firmly rooted on individual behaviors of the agents. However since it reproduces what is seen on the market, the way market participants act has to be consistent with the model. Therefore, as explained in the rest of this section and in Section \ref{theo}, ex post, an agent based interpretation of such a statistical model can still be given.

\subsection{Description of the model}

The heuristic of the model is very simple. When the bid-ask is given, market takers know the price for which they can buy and the price for which they can sell. However, they have their own opinion about the ``fair" price of the asset, inferred from all available market data and their personal views. In the latter, we assume that there exists an efficient price, representing this opinion. Of course this efficient price should not be seen as an ``economic price'' of the asset, but rather as a market consensus at a given time about the asset value. The idea of the model with uncertainty zones is that for large tick assets, at a given time, the difference between the efficient price and the best accessible price on the market for buying (resp. selling) is sometimes too large so that a buy (resp. sell) market order can occur.

\subsubsection{Efficient price}

We propose here a simplified version of the model with uncertainty zones, see \cite{rr1} for a more general version. The first assumption on the model is the following:
\begin{itemize}
\item[H$1$] {\it There is a latent efficient price with value $X_t$ at time $t$, which is a continuous semi-martingale of the form
$$X_t=X_0+\int_0^t a_u \text{d}u+\int_0^t \sigma_{u-} \text{d} W_u,$$
where $W_t$ is a $\mathcal{F}_t$-Brownian motion, with $\mathcal{F}$ a filtration for which $a_u$ is progressively measurable and locally bounded and $\sigma_u$ is an adapted right continuous left limited process.}
\end{itemize}
Following in particular the works by A\"it-Sahalia {\it et al.}, see \cite{amz,zma}, using such kind of efficient price process when building a microstructure model has become very popular in the recent financial econometrics literature. Indeed, it enables to easily retrieve standard Brownian type dynamics in the low frequencies, which is in agreement with both the behavior of the data and the classical mathematical finance theory. Also, our assumptions on the efficient price process are very weak, allowing in particular for any kind of time varying or stochastic volatility. Of course this efficient price is not directly observed by market participants. However, they may have their own opinion about its value.

\subsubsection{Uncertainty zones and dynamics of the last traded price}

Let $\alpha$ be the tick value of the asset. We define the uncertainty zones as bands around the mid tick values with width $2\eta\alpha$, with $0<\eta\leq 1$ a given parameter. The dynamics of the last traded price, denoted by $P_t$, is obtained as a functional of the efficient price and the uncertainty zones. Indeed, in order to change the transaction price, we consider that market takers have to be ``convinced'' that it is reasonable, meaning that the efficient price must be close enough to a new potential transaction price. This is translated in Assumption H$2$.

\begin{itemize}
\item[H$2$]  {\it Let $t_0$ be any given time and $P_{t_0}$ the associated last traded price value. Let $\tau_{t_0}^u$ be the first time after $t_0$ where $X_t$ upcrosses the uncertainty zone above $P_{t_0}$, that is hits the value $P_{t_0}+\alpha/2+\eta\alpha$. Let $\tau_{t_0}^d$ be the first time after $t_0$ where $X_t$ downcrosses the uncertainty zone below $P_{t_0}$, that is hits the value $P_{t_0}-\alpha/2-\eta\alpha$. Then, one cannot have a transaction at some time $t>t_0$ at a price
strictly higher (resp. smaller) than $P_{t_0}$ before $\tau_{t_0}^u$ (resp. $\tau_{t_0}^d$). Moreover, if $\tau_{t_0}^u<\tau_{t_0}^d$ (resp. $\tau_{t_0}^u>\tau_{t_0}^d$) one does have a transaction at the new price $P_{t_0}+\alpha$ (resp. $P_{t_0}-\alpha$) at time $\tau_{t_0}^u$ (resp. $\tau_{t_0}^d$).}
\end{itemize}

In fact, when associating it to Assumption H$3$, we will see that Assumption H$2$ can be understood as follows: at any given time, a buy (resp. sell) market order cannot occur if the current value of the efficient price is too far from the best ask (resp. bid).

Remark that Assumption H$2$ implies that the transaction price only jumps by one tick, which is fairly reasonable for large tick assets. However, imposing jumps of only one tick and that a transaction occurs exactly at the times the efficient price exits an uncertainty zone is done only for technical convenience. Indeed, it can be easily relaxed in the setting of the model with uncertainty zones, see \cite{rr1}.

\begin{figure}
\centering
\includegraphics[width=1\textwidth]{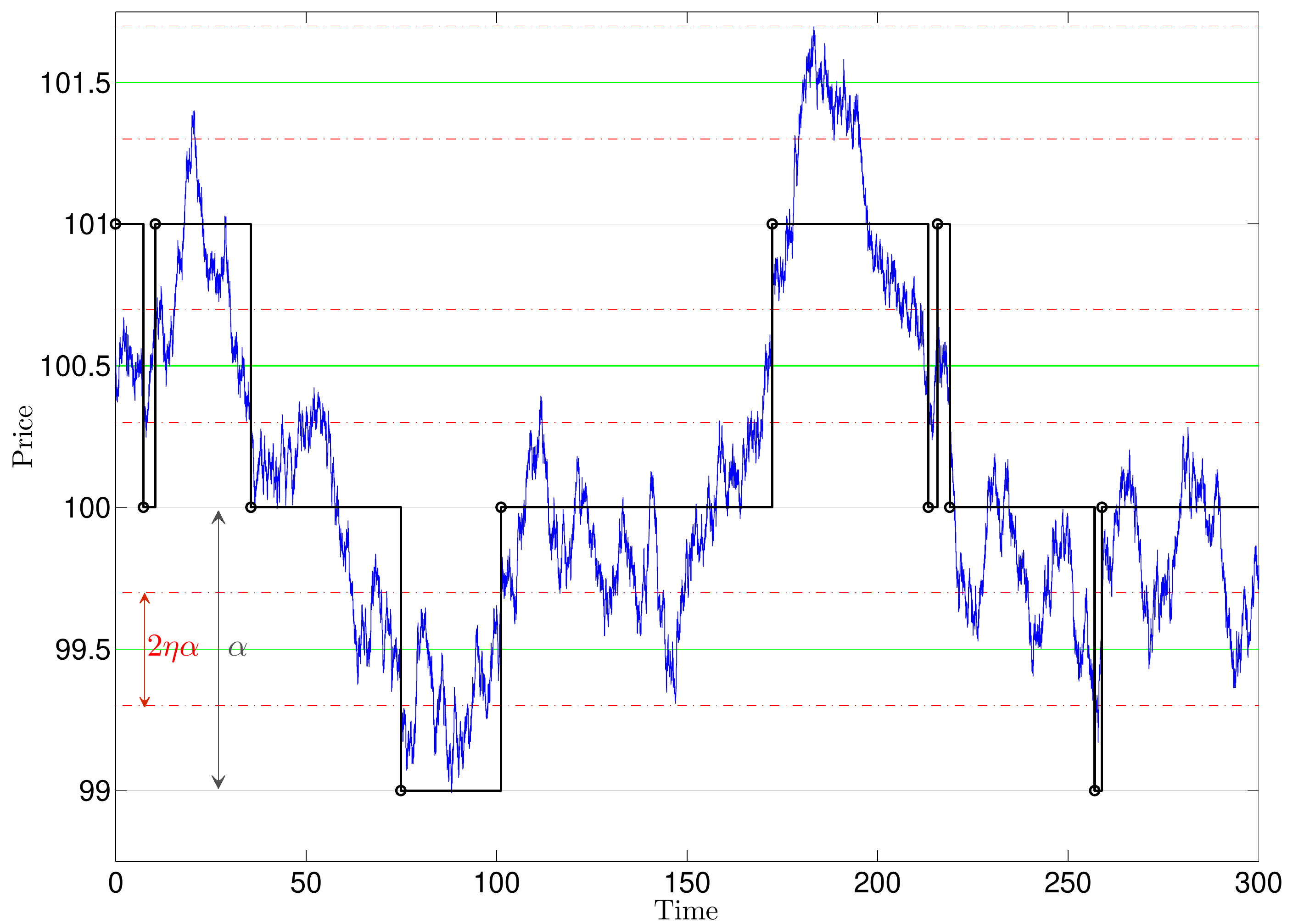}
\caption{The model with uncertainty zones: example of sample path. The efficient price is drawn in blue. The light gray lines drawn at integers form the tick grid of width $\alpha$. The red dotted lines are the limits of the uncertainty zones of width $2\eta\alpha$. Finally the last traded price is the black stepwise curve. The circles indicate a change in the price when the efficient price crosses an uncertainty zone.}\label{fig:Le_graphenodelay2}
\end{figure}

A sample path of the last traded price in the model with uncertainty zones is given in Figure \ref{fig:Le_graphenodelay2}.

\subsubsection{Bid-ask spread}

In this work, we focus on large tick assets. By this we mean assets whose bid-ask spread is essentially constant and equal to one tick. Therefore we make the following assumption in the model.
\begin{itemize}
\item[H$3$] {\it The bid-ask spread is constant, equal to the tick value $\alpha$.}
\end{itemize}
In practice, the preceding assumption means that if at some given time the spread is not equal to one tick, limit orders immediately fill the gap. Remark that we do not impose the efficient price to lie inside the bid-ask quotes. However, the dynamics of the bid-ask quotes still need to be compatible with Assumption H$2$.

\begin{figure}
\centering
\includegraphics[width=\textwidth]{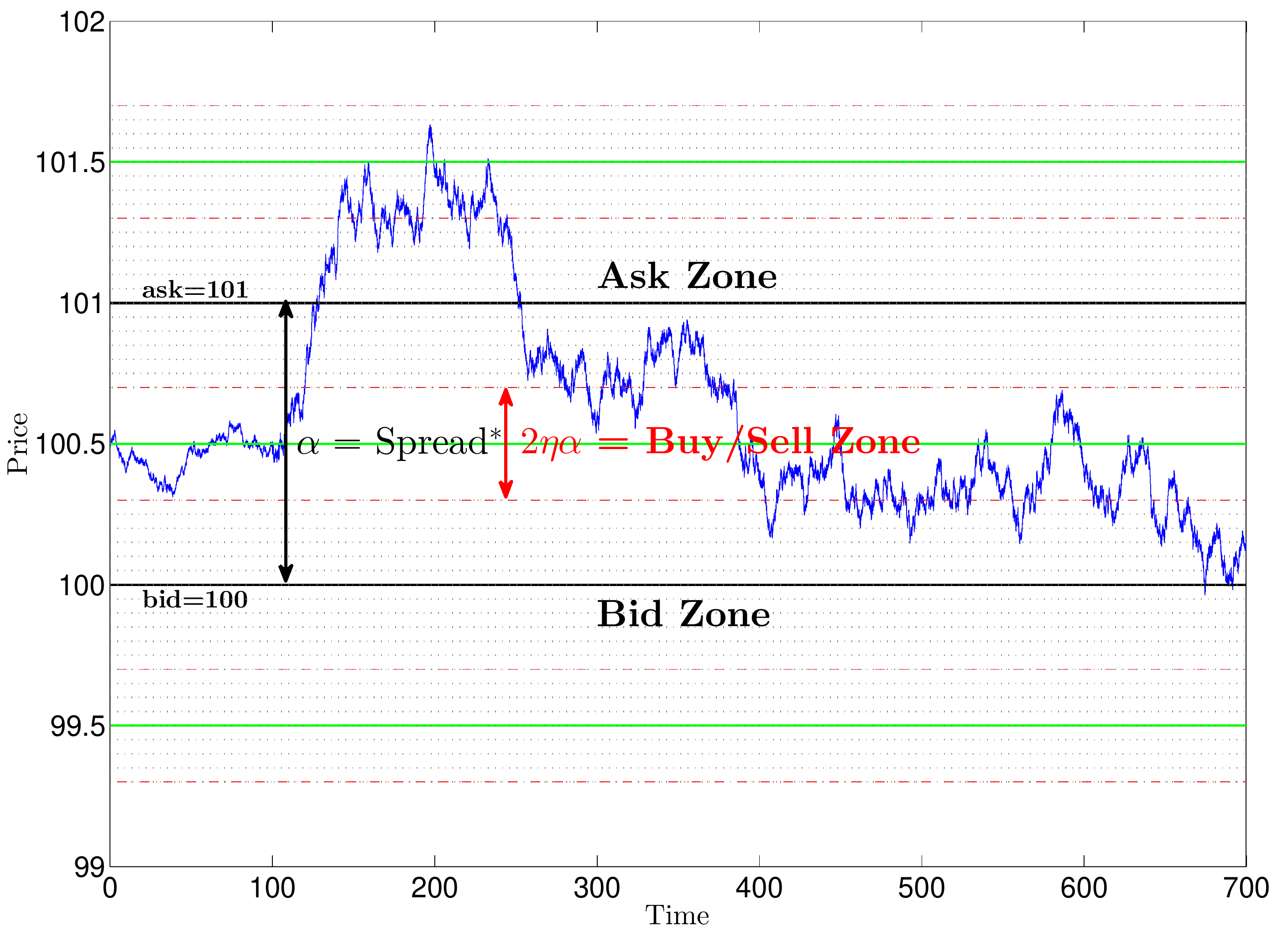}
\caption{The three different zones when the bid-ask is 100-101 and the tick value is equal to one. The red dotted lines are the limits of the uncertainty zones. The uncertainty zone inside the spread is the buy/sell zone. The upper dotted area is the ask zone and the lower dotted area is the bid zone.}\label{fig:Uncertainty_Shade3}
\end{figure}

Within bid-ask quotes of the form $[b,b+\alpha]$, the width of the uncertainty zone represents the range of values for $X_t$ where  transactions at the best bid and the best ask can both occur. The size of this range is $2\eta\alpha$. Therefore, it is natural to view the quantity $2\eta\alpha$ as an {\it implicit spread}, see Section \ref{num}. More precisely, for given bid-ask quotes $[b,b+\alpha]$, Assumptions H$2$ and H$3$ enable us to define three areas for the value of the efficient price process $X_t$:

\begin{itemize}
\item The bid zone: $(b-\alpha/2-\eta\alpha,b+\alpha/2-\eta\alpha)$, where only sell market orders can occur.
\item The buy/sell zone: $[b+\alpha/2-\eta\alpha,b+\alpha/2+\eta\alpha]$, where both buy and sell market orders can occur. It coincides
with the uncertainty zone.
\item The ask zone: $(b+\alpha/2+\eta\alpha,b+3\alpha/2+\eta\alpha)$, where only buy market orders can occur.
\end{itemize}
This is summarized in Figure \ref{fig:Uncertainty_Shade3}.


\subsection{Comments on the model and the parameter $\eta$}

\paragraph{Reproducing and quantifying microstructure effects}

\begin{itemize}
\item The model is particularly parsimonious since it only relies on an efficient price process and the uncertainty zones parameter $\eta$. Despite its simplicity, this simple model accurately reproduces all the main stylized facts of market data, see \cite{rr1}.
\item The parameter $\eta$ turns out to measure the intensity of microstructure effects. Indeed, all the microstructure phenomena such as the autocorrelations of the tick by tick returns or the law of the durations between price changes can be easily quantified through the single parameter $\eta$, see again \cite{rr1}. For example, let us consider the case where the volatility process is constant equal to $\sigma$. Then it is shown in \cite{rr2} that as $\alpha$ goes to zero,
\begin{equation}\label{vol1}\sum_{0\leq t_i< t_{i+1}\leq t}(P_{t_{i+1}}-P_{t_{i}})^2\rightarrow \frac{\sigma^2t}{2\eta},\end{equation} where the $t_i$'s denote the transaction times with price change. Therefore, if $\eta<1/2$, we recover here the very well known stylized fact that the high frequency realized variance of the observed price is larger than those of the efficient price, which is $\sigma^2t$. More precisely, in that case, we obtain a decreasing behavior of the so called signature plot, that is the function from $\mathbb{N^*}$ to $\mathbb{R}^+$ defined by
\begin{equation}\label{sigplot}
\Delta\rightarrow \sum_{i=1}^{\lfloor nt/\Delta\rfloor}\big(P_{\Delta i/n}-P_{\Delta(i-1)/n}\big)^2,
\end{equation}
where $n$ is a fixed ultra high frequency sampling value for the last traded price. Since the seminal paper \cite{abdl}, this is considered in the econometric literature as one of the most distinctive features of high frequency data. In fact, the estimated values of $\eta$ are indeed systematically found to be smaller than $1/2$. In our framework, this can be nicely explained from a theoretical point of view, see Section \ref{theo}.



\item When the tick size is large, market participants are not indifferent to a one tick price change and the traded price is modified only if market takers are convinced it is reasonable to change it. This is exactly translated in our model through the key parameter $\eta$. Indeed, in order to have a new transaction price, $X_t$ needs to reach a barrier which is at a distance $\eta\alpha$ from the mid tick. So, when $\eta$ is small, a very small percentage of the tick value is considered enough for a price change, meaning the tick value is very large and conversely. A different point of view is to consider that market participants have a certain resolution, or precision at which they infer the efficient price $X_t$. This resolution is quantified by $\eta$, and is close to the tick value when $\eta$ is close to $1/2$.

\item The width of a buy/sell zone is $2\eta\alpha$. Thus, if $\eta$ is small, there is a lot of mean reversion in the price and the buy/sell zones are very small: the tick size is very large. If $\eta$ is close to $1/2$, the last traded price can be seen as a sampled Brownian motion, there is no microstructure effects, and the width of the buy/sell zones is one tick: the tick size is, in some sense, {\it optimal}, see Section \ref{change}.

\item In fact, we can give a much more precise interpretation of $\eta$. Indeed, we show in the next section that the quantity $2\eta\alpha$ can be seen as an implicit spread. A by product of this is the fact that
$\eta$ can indeed be viewed as a suitable measure for the tick size.

\end{itemize}

\paragraph{Statistical estimation of $\eta$ and of the volatility}

\begin{itemize}

\item The parameter $\eta$ can be very easily estimated as follows.  We define an alternation (resp. continuation) of one tick
as a price jump of one tick whose direction is opposite
to (resp. the same as) the one of the preceding price jump. Let
$N_{\alpha ,t}^{( a) }$ and $N_{\alpha ,t}^{( c) }$ be
respectively the number of alternations and continuations of one
tick over the period $[0,t]$.
It is proved in \cite{rr2} that as the tick value goes to zero, a consistent estimator of $\eta$ over $[0,t]$ is given by
$$
\widehat{\eta}_{\alpha ,t}=\frac{N_{\alpha ,t}^{( c)
}}{ 2N_{\alpha ,t}^{( a) }}.
$$

\item The model with uncertainty zones enables to retrieve the value of the efficient price at the time
$t_i$ of the $i$-th price change by the simple relation
\begin{equation*}
X_{t_{i}}=P_{t_{i}}-sign(P_{t _{i}}-P_{t _{i-1}})(1/2-\eta
)\alpha.
\end{equation*}%
Hence, since we can estimate $\eta$, we can recover
$X_{t _{i}}$ from $P_{t_{i-1}}$ and $P_{t_{i}}$. This is very convenient for building
statistical procedures relative to the efficient. For example, the realized variance computed over
the estimated values of the efficient price between $0$ and $t$:
\begin{equation}\label{estimvol}
\widehat{\sigma^2_{[0,t]}}=\sum_{t_i\leq t}\big(\widehat{X}_{t_{i}} -\widehat{X}_{t
_{i-1}}\big) ^{2},
\end{equation}%
where  $\widehat{X}_{t_{i}}=P_{t_{i}}-sign(P_{t _{i}}-P_{t
_{i-1}})(1/2-\hat{\eta}_{\alpha,t})\alpha,$ is a very sharp estimator
of the integrated variance of the efficient price over $[0,t]$:
$$\int_0^t\sigma^2_u\text{d}u.$$
The accuracy of this
estimator is $\alpha$ and its asymptotic theory is available in
\cite{rr2}.\\

\end{itemize}

\section{Implicit spread and volatility per trade: empirical study}\label{num}

A buy/sell zone $[b+\alpha/2-\eta\alpha,b+\alpha/2+\eta\alpha]$ is a kind of a frontier, such that crossing it makes market takers change their view on the efficient price. It is a sort of tolerance area defined by their risk aversion to losing one tick. The width of this zone, $2\eta\alpha$, also corresponds to the size of the (efficient) price interval for which market takers are both ready to buy and to sell. This is why we see it as a kind of a spread: \textit{the market taker's implicit spread}. In view of this interpretation, we consider the similarities in the properties of this implicit spread to those of the conventional spread. In particular, we look at the spread-volatility relationship described in Section \ref{intro} that stipulates that the spread is generally proportional to the volatility per trade. In this section, we empirically verify this relationship using our implicit spread and see that it holds remarkably well.

This approach follows in the some sense those of Roll in \cite{rol}. In this paper, the author addresses the problem of estimating the bid-ask spread if one has only access to transaction data. He shows that in his framework, the quantity $\sqrt{-2 \text{Cov}}$, where Cov denotes the first order autocovariance of the price increments, is a good proxy for the spread. This is particularly interesting since this autocovariance can be expressed in term of $\eta$. Indeed, in the model with uncertainty zones, we have
$$\sqrt{-2\text{Cov}}=\sqrt{\frac{2-4\eta}{1+2\eta}}\alpha.$$
Thus the link between a parameter such as $\eta$ and a kind of spread is already present in \cite{rol}.
However, in \cite{rol}, the author works at a completely different time scale and
this measure is not relevant for large tick assets traded at high frequency on electronic markets. In particular, it decreases
with $\eta$ for $\eta$ between $0$ and $1/2$, which is not consistent with the empirical results.

\subsection{Definition of the variables}
In this section, we want to investigate the relationship
$$\text{Implicit spread}\sim\text{Volatility per trade}+\text{constant}.$$
The implicit spread and the volatility per trade are computed on a daily basis.
Following the approach of Madhavan {\it et al.} \cite{mrr}, the volatility over the period, denoted by $\sigma$, is taken with reference to the efficient price. We use the estimator $\widehat{\sigma^2_{[0,t]}}$ of the cumulative variance of the efficient price over $[0,t]$ introduced in Equation \eqref{estimvol} (renormalized in square of currency unit) and set
$$\widehat\sigma=\sqrt{\widehat{\sigma^2_{[0,t]}}}.$$
Then we define the volatility per trade by
$\widehat\sigma/\sqrt{M},$ where $M$ denotes the total number of trades (all the transactions, changing the last traded price or not) over $[0,t]$. From now on, abusing notation slightly, we make no difference between the parameters and their estimators.
Therefore, our relationship can be rewritten
$$\eta\alpha\sim\frac{\sigma}{\sqrt{M}}+\phi.$$
In the sequel we also need to compute an average daily spread, denoted by $S$, which is in practice not exactly equal to one. This spread is measured as the average over the considered time period of the observed spreads right before the trades. Thus, for each asset, we record everyday the vector $(\eta\alpha,\sigma,M,S)$.

\subsection{Description of the data}

We restrict our analysis to assets traded in well regulated electronic markets which match the framework of the electronic double auction. We use data of 10 futures contracts on assets of different classes and traded in different exchanges. The database\footnote{Data provided by QuantHouse. http://www.quanthouse.com/} has millisecond accuracy and was recorded from 2009, May 15 to 2009, December 31.

On the CBOT exchange, we use the 5-Year U.S. Treasury Note Futures (BUS5) and the futures on the Dow Jones index (DJ). On the CME, we use the forex EUR/USD futures (EURO) and the futures on the SP500 index (SP). On the EUREX exchange, we use three interest rates futures based on German government debt: The 10-years Euro-Bund (Bund), the 5-years Euro-Bobl (Bobl) and the 2-years Euro-Schatz (Schatz). Note that the tick value of the Bobl changed on 2009, June 15. Thus, we write Bobl 1 when referring to the Bobl before this date and Bobl 2 after it. We also investigate futures on the DAX index (DAX) and on the EURO-STOXX 50 index (ESX).
Finally we use the Light Sweet Crude Oil Futures (CL) traded on the NYMEX. As for their asset classes, the DJ, SP, DAX and ESX are equity futures, the BUS5, Bund, Bobl and Schatz are fixed income futures, the EURO is a foreign exchange rate futures and finally the CL is an energy futures. On the exchanges, the settlement dates for these future contracts are standardized, one every three months (March, June, September and December) and generally three future settlement months are trading at the same time. We deal with this issue by keeping, on each day, the contract that recorded the highest number of trades and discarding the other maturities.

These assets are all large tick assets, with a spread almost always equal to one tick. To quantify this, for each asset, we compute everyday the percentage of trades for which the value of the observed spread right before the trade is equal to one tick. The average of these values is denoted by \#$S_=$ and is reported in Table \ref{tab:intro:Stats1}, together with other information about the assets, notably the average values of $\eta$, denoted by \#$\eta$.
\begin{table}[h!]
\begin{small}
\centering
\begin{tabular}{rrrrccrr}
\toprule
\multicolumn{1}{c}{Futures} & \multicolumn{1}{c}{Exchange} & \multicolumn{1}{c}{Class} & \multicolumn{1}{c}{Tick Value} & Session & \multicolumn{1}{c}{\# Trades/Day} & \multicolumn{1}{c}{\#$\eta$} & \multicolumn{1}{c}{\#$S_=$} \\
\midrule
\multicolumn{1}{l}{BUS5} & \multicolumn{1}{l}{CBOT} & \multicolumn{1}{l}{Interest Rate} & 7.8125 \$& 7:20-14:00 & 26914 & 0.233 & 94.9 \\
\multicolumn{1}{l}{DJ} & \multicolumn{1}{l}{CBOT} & \multicolumn{1}{l}{Equity} & 5.00 \$& 8:30-15:15 & 48922 & 0.246 & 81.8 \\
\multicolumn{1}{l}{EURO} & \multicolumn{1}{l}{CME} & \multicolumn{1}{l}{FX} & 12.50 \$& 7:20-14:00 & 46520 & 0.242 & 90.6 \\
\multicolumn{1}{l}{SP} & \multicolumn{1}{l}{CME} & \multicolumn{1}{l}{Equity} & 12.50 \$& 8:30-15:15 & 118530 & 0.035 & 99.6 \\\hline
\multicolumn{1}{l}{Bobl 1} & \multicolumn{1}{l}{EUREX} & \multicolumn{1}{l}{Interest Rate} & \euros{5.00} & 8:00-17:15 & 18531 & 0.268 & 95.3 \\
\multicolumn{1}{l}{Bobl 2} & \multicolumn{1}{l}{EUREX} & \multicolumn{1}{l}{Interest Rate} & \euros{10.00} & 8:00-17:15 & 11637 & 0.142 & 99.2 \\
\multicolumn{1}{l}{Bund} & \multicolumn{1}{l}{EUREX} & \multicolumn{1}{l}{Interest Rate} & \euros{10.00} & 8:00-17:15 & 25182 & 0.138 & 98.1 \\
\multicolumn{1}{l}{DAX} & \multicolumn{1}{l}{EUREX} & \multicolumn{1}{l}{Equity} & \euros{12.50} & 8:00-17:30 & 39573 & 0.275 & 72.7 \\\hline
\multicolumn{1}{l}{ESX} & \multicolumn{1}{l}{EUREX} & \multicolumn{1}{l}{Equity} & \euros{10.00} & 8:00-17:30 & 35121 & 0.087 & 99.5 \\
\multicolumn{1}{l}{Schatz} & \multicolumn{1}{l}{EUREX} & \multicolumn{1}{l}{Interest Rate} & \euros{5.00} & 8:00-17:15 & 9642 & 0.122 & 99.4 \\
\multicolumn{1}{l}{CL} & \multicolumn{1}{l}{NYMEX} & \multicolumn{1}{l}{Energy} & 10.00 \$& 8:00-13:30 & 73080 & 0.228 & 75.7 \\
\bottomrule
\end{tabular}%
\caption{Data Statistics. The \emph{Session} column indicates the considered trading hours (local time). The sessions are chosen so that we get enough liquidity and are not the actual sessions.}\label{tab:intro:Stats1}
\end{small}
\end{table}

\subsection{Graphical analysis}

\begin{figure}
\centering
\includegraphics[width=\textwidth]{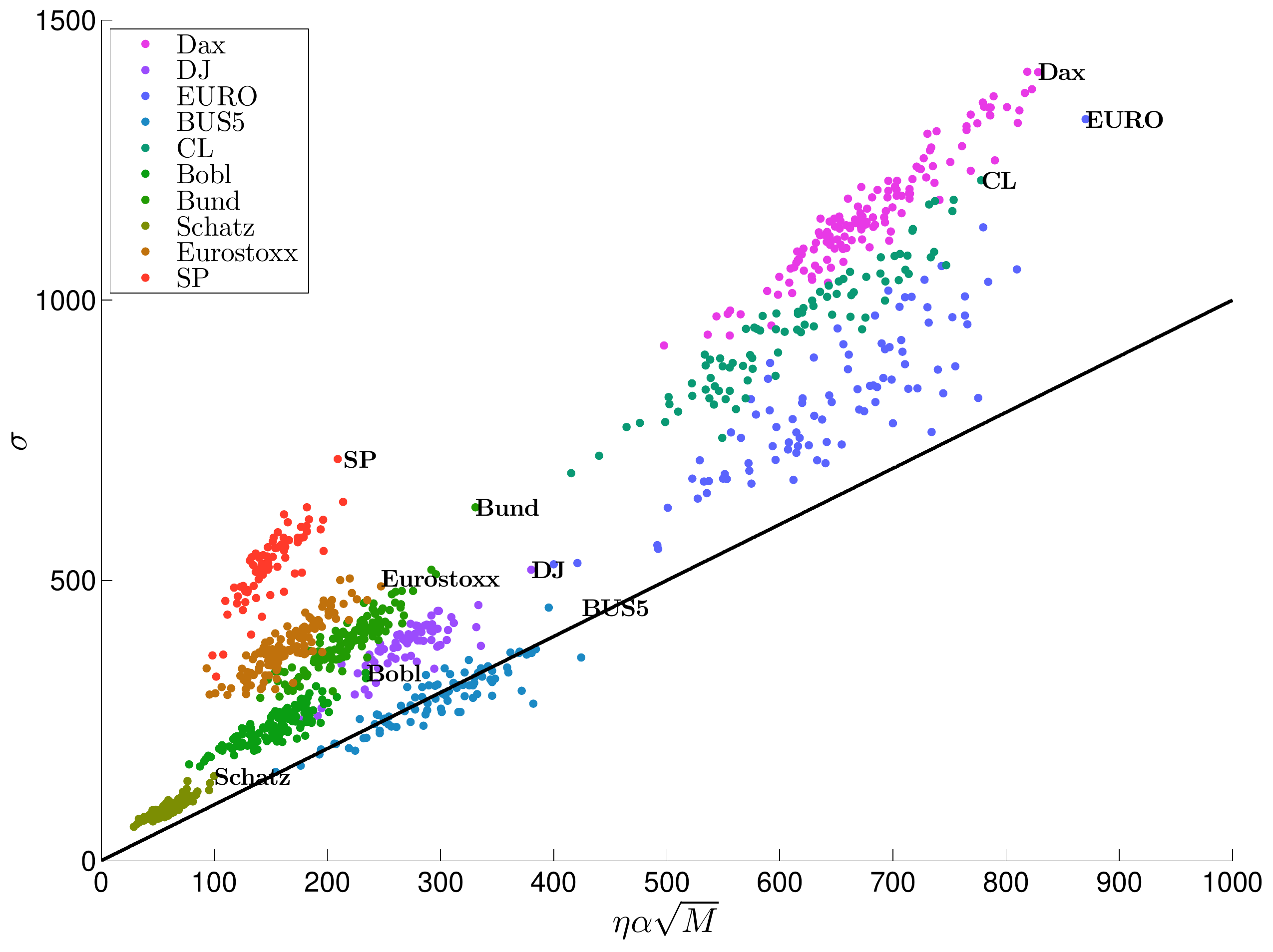}
\caption{Cloud $(\eta\alpha\sqrt{M},\sigma)$. The black line is the line $y=x$.}\label{fig:cloud}
\end{figure}

In order to have a first idea of the relevance of our implicit spread, we give the cloud $(\eta\alpha\sqrt{M},\sigma)$ in Figure \ref{fig:cloud}\footnote{We give the cloud $(\eta\alpha\sqrt{M},\sigma)$ rather than $(\eta\alpha,\sigma/\sqrt{M})$ in order to get more readable values.}. Each point represents one asset, one day.

The results are quite striking. At the visual level, we obtain a linear relationship between $\sigma$ and $\eta\alpha\sqrt{M}$, with the same slope for each asset but different intercepts. In particular, Figure \ref{fig:cloud} looks very similar to the kind of graph obtained in \cite{wbkpv}, where the real spread is used with small tick assets.

\subsection{Linear regression}

In order to get a deeper analysis of the relationship between the implicit spread and the volatility per trade, we consider the linear regression associated to the relation

$$\eta\alpha\sim\frac{\sigma}{\sqrt{M}}+\phi.$$

The constant $\phi$ includes costs and profits related to the inventory control and to the fact that the average spread of the assets is not exactly equal to one tick. In the spirit of the approaches mentioned in Section \ref{intro}, this last fact should imply a slightly larger market makers profit than in the case where the spread is exactly equal to one tick. Therefore, we consider that for each asset, the constant $\phi$ is proportional to the average daily spread. Thus we consider the daily regression with unknown $p_1,p_2,p_3$:
\begin{equation}\label{reg}
\sigma=p_1\eta\alpha\sqrt{M}+p_2S\sqrt{M}+p_3.
\end{equation}
The results are given in Table \ref{tablereg}\footnote{Note that we have merged the data corresponding to Bobl 1 and Bobl 2. Anyhow, the regression parameters are very close when considering them separately.}.

\begin{table}[h!]
\begin{small}
\centering
\begin{tabular}{lcccc}
\toprule
Asset &\textbf{$p_1$}&\textbf{$p_2$}&\textbf{$p_3$}&\textbf{$R^2$}\\
\midrule
BUS5    &0.67 [0.55,0.79]    &0.10 [0.06,0.14]    &-40.21 [-76.28,-4.14]    &0.84\\
DJ      &0.93 [0.71,1.15]    &0.07 [0.01,0.13]    &38.90 [-18.19,96.00]     &0.73\\
EURO    &1.31 [1.11,1.51]    &0.02 [-0.02,0.07]   &-89.23 [-211.08,32.62]   &0.75\\
SP      &1.67 [1.37,1.96]    &0.07 [0.05,0.08]    &-2.84  [-69.90, 64.21]   &0.83\\\hline
Bobl    &0.91 [0.84,0.97]    &0.08 [0.07,0.09]    &19.04 [4.41,33.67]       &0.90\\
Bund    &1.11 [1.01,1.20]    &0.11 [0.09,0.13]    &-29.99 [-54.16,-5.82]    &0.92\\
Dax     &1.09 [1.01,1.16]    &0.11 [0.10,0.13]    &54.94 [23.02,86.86]      &0.97\\
ESX     &0.89 [0.78,1.01]    &0.13 [0.11,0.15]    &-10.15 [-37.71,17.41]    &0.90\\\hline
Schatz  &0.80 [0.71,0.90]    &0.10 [0.07,0.12]    &-0.93 [-9.78,7.92]       &0.88\\
CL      &0.97 [0.89,1.05]    &0.11 [0.09,0.12]    &-11.14 [-51.20,28.92]    &0.97\\
\bottomrule
\end{tabular}
\caption{Estimation of the linear model with 95\% confidence intervals.}\label{tablereg}
\end{small}
\end{table}

\begin{figure}
\centering
\includegraphics[width=\textwidth]{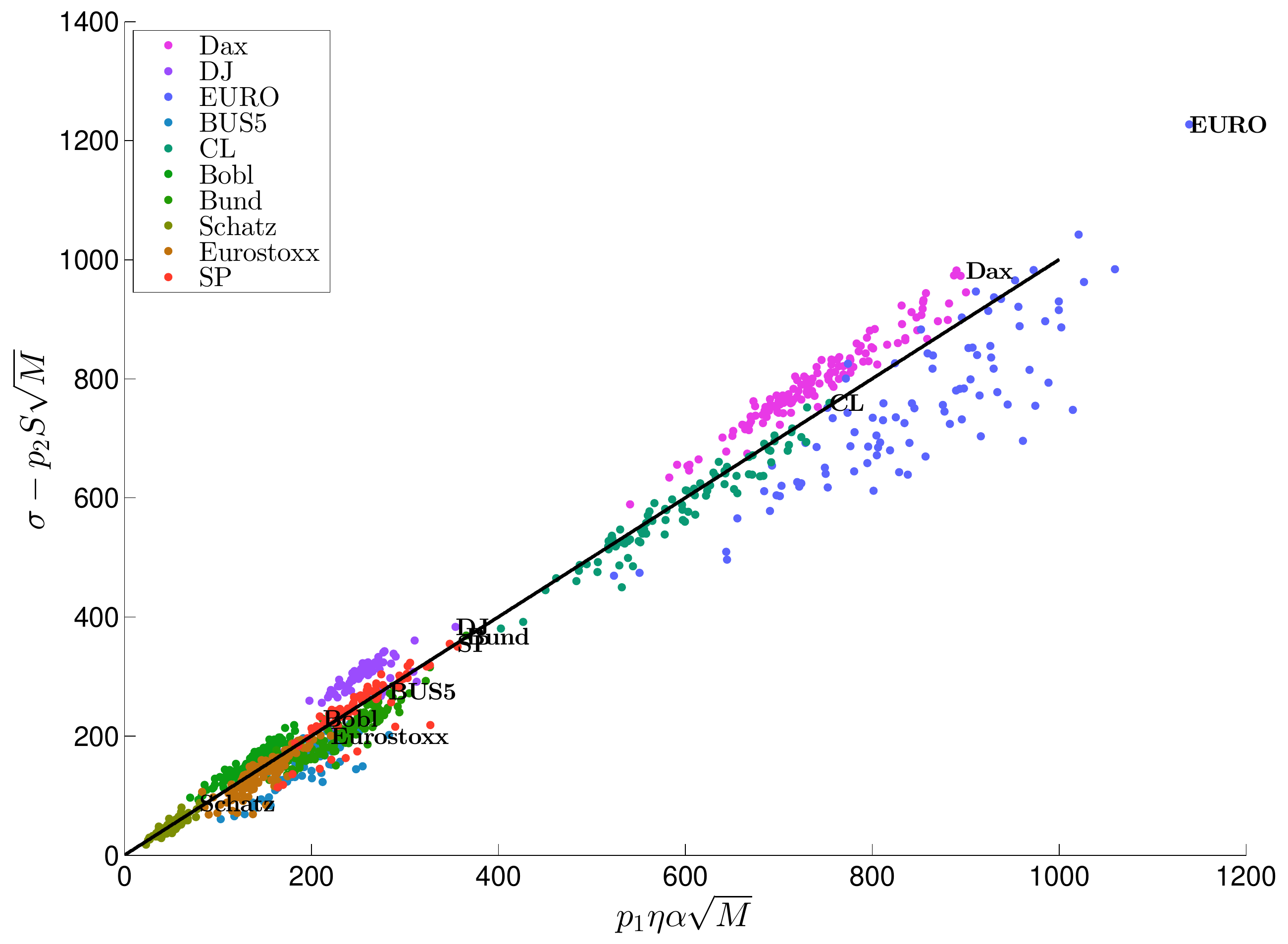}
\caption{Cloud $(p_1\eta\alpha\sqrt{M},\sigma-p_2S\sqrt{M})$. The black line is the line $y=x$.}\label{fig:cloud2}
\end{figure}

By looking at the $R^2$ statistics, we can notice that the linear fits are very good. More interestingly, we see that the values of $p_1$ are systematically very close to 1. We explain this from a theoretical point of view in the next section. Surprisingly enough, we also remark that the constant $p_2$ has the same order of magnitude for all the assets (about 0.1). Finally, in order to show that the parameter $p_3$ is negligible, the cloud $(p_1\eta\alpha\sqrt{M},\sigma-p_2S\sqrt{M})$ is given in Figure \ref{fig:cloud2}. On this figure, all the points are indeed very close to the line $y=x$.

\section{Implicit spread and volatility per trade: a simple equilibrium model}\label{theo}

In our approach, the relationship between the implicit spread $\eta\alpha$ and the volatility per trade can be theoretically justified in a very natural way. Indeed, we use a simple equilibrium equation between profits and losses of market makers and market takers. To do so, in the spirit of Madhavan {\it el al.} \cite{mrr}, we compute the ex post expected cost relative to the efficient price of a market order.

\subsection{A profit and loss equality}

To fix ideas, let us consider a market order leading to an upward price change at time $t$. We write $X_t$ for the efficient price at the transaction time $t$ and assume it lies inside the bid-ask quotes at the transaction time. Therefore, from the model, the market order has been done at price $P_t=X_t+\alpha/2-\eta\alpha$. After this transaction, there are two possibilities:

\begin{itemize}

\item The next move of the transaction price is a downward move at price $P_t-\alpha$: it means the efficient price has crossed the barrier $X_t-2\eta\alpha$. This occurs with probability $\frac{1}{1+2\eta}$.
\item The next move of the transaction price is an upward move at price $P_t+\alpha$: it means the efficient price has crossed the barrier $X_t+\alpha$. This occurs with probability $\frac{2\eta}{1+2\eta}$.
\end{itemize}

Therefore, the ex post expected profit and loss of such a market order is
$$(X_t+\alpha/2-\eta\alpha)-\Big(\frac{1}{1+2\eta}(X_t-2\eta\alpha)+\frac{2\eta}{1+2\eta}(X_t+\alpha)\Big)=\alpha/2-\eta\alpha.$$
Thus, contrary to the classical efficiency condition of small tick markets which states that the ex post expected cost of a market order should be zero, see for example \cite{wbkpv}, in the large tick case, it is positive. This means that confronted with this large tick, market takers are ready to lose $\alpha/2-\eta\alpha$ in order to obtain liquidity.

As already seen, following Wyart {\it et al.} \cite{wbkpv}, the average profit and loss per trade per unit of volume of the market makers is well approximated by $\alpha/2-\sigma/\sqrt M$. The gain of the market makers being the loss of the market takers, this leads to
$$\eta\alpha\sim\sigma/\sqrt M.$$
Thus, using a theoretical approach inspired by Madhavan {\it et al.} \cite{mrr} and Wyart {\it et al.} \cite{wbkpv}, we can explain our empirical finding that $\eta\alpha$ plays the role of spread for large tick assets.

\subsection{A simple agent-based explanation of microstructure effects}

A distinctive feature of high frequency data, particularly of large tick assets, is the decreasing behavior of the signature plot \eqref{sigplot}. Many statistical models aim at reproducing this decreasing shape. Most commonly, they use a measurement error approach for the price (microstructure noise models), see among others \cite{amz,abdl,br,hl,r,zma}. A pleasant consequence of our approach is that it provides an agent based explanation of a phenomenon mostly viewed as a statistical stylized fact.

Recall that the ex post expected cost of a market order is $\alpha/2-\eta\alpha$. This does explain why for large tick assets with average spread close to one tick, the parameter $\eta$ is systematically smaller than $1/2$, which means the signature plot is decreasing. Otherwise we would be in a situation where the cost of market orders is negative and market makers lose money. To avoid that, market makers would naturally increase the spread, what they can always do.

\section{Changing the tick value}\label{change}

Market designers face the question of choosing a tick value. This is an extremely important and sensitive issue, especially because of its impact on high frequency trading. Surprisingly, it has not been given much consideration in the quantitative academic literature. Our approach seems then to be one of the first attempt to fill this gap.

Fixing the tick value is an intricate problem, see for example \cite{har2,har3}. On the one hand, when the tick value is too small, one tick is not really significant, neither for market makers nor for market takers. Therefore, it is very complicated for market makers to choose levels where they should fix their quotes. Furthermore, the order books are very unstable since market participants do not hesitate changing marginally the price of their limit orders in order to gain in priority, which can be very discouraging for market makers. In particular, market participants having only access to a few lines of the order book (typically five or ten), if these lines are not reliable or only provide vanishing liquidity, they may not be able to assess the prices. On the other hand, it is clear that a tick value which is too large prevents the price from moving freely according to the views of market participants whose valuation accuracy for the asset is smaller than one tick.

If the tick value is not satisfying, market platforms have the possibility to change it. Such a modification implies changes in various market quantities (number of trades, spread, liquidity,...). The first thing the platform designer needs to do is to understand the desired effects of this change of tick. This is already a difficult question, see Section \ref{optick}. Even in the case where market designers have a clear idea of the situation they want to reach, they still face the problem of the way to reach it. Indeed, it is commonly acknowledged that tick values have to be determined by trial and error and that the success of a change in the tick value can only be assessed ex post, on the basis of the obtained effects. Thus, only few predictive models have been designed in the literature, see for example \cite{har1}, and the consequences of a change in the tick value have been essentially studied from an empirical point of view, see \cite{ahn,bac,bes,bour,chu,chu2,gol,gri,lau,ron,wu,zha}.

\subsection{The effects of a change in the tick value}

We assume we are dealing with a large tick asset. In that case, our approach enables us to forecast {\it ex ante} the consequences of a change in the tick value on some market quantities, in particular $\eta$ which is the parameter quantifying the intensity of microstructure effects.

In the following analysis, the variables of interest are: the tick value $\alpha$, the daily volatility $\sigma$, the daily number of transactions $M$, the traded volume within the day $V$, the regression estimates $p_1$ and $p_2$, and the parameter $\eta$. We put an index $0$ for denoting a variable before the change in the tick value and no index for the same variable after the change in the tick value. We discuss in more details these seven variable in the following:

\begin{itemize}
\item $\alpha_0$ and $\alpha$ are fixed by the exchange.
\item The volatility and the daily traded volume are macroscopic, fundamental quantities and should remain essentially invariant after a change in the tick value. Therefore we will assume:
$$\sigma=\sigma_0,~~V=V_0.$$
\item The regression parameters $p_{1,0}$ and $p_{2,0}$ are known. The parameters $p_{1}$ and $p_{2}$ are a priori unknown but we have shown that in practice $p_1$ is systematically close to $1$ and $p_2$ is small, close to $0.1$. Moreover, this value for $p_1$ is clearly explained from a theoretical point of view. Thus we will assume:
$$p_1\sim p_{1,0}\sim 1\text{ and }p_2\sim p_{2,0},~p_2\in[0,0.1].$$
\item The daily number of trades $M$ should not be an invariant quantity. Assume for example that at any time, the average cumulative latent liquidity available up to price $p$ is of the form $f(|p-midprice|)$, with $f$ an increasing function. We assume also that each market taker takes a fixed proportion of the liquidity at the top of the book (equal to $f(\alpha/2)$). Then, when the tick size decreases, the available liquidity at the best levels also decreases. The daily traded volume being approximately constant, this implies an increase in the number of transactions $M$. We will use such a function $f$ with a reasonable shape, more precisely:
$$f(x)=c*x^{\beta}, \beta>0,$$
focusing on the classical cases $\beta=1$ (linear case) and $\beta=1/2$ (square root concave case).
\item The parameter $\eta_0$ is known, but the parameter $\eta$ is a priori unknown.
\end{itemize}

Recall that we consider a large tick asset whose tick value is modified. We assume we remain in a regime where the spread is approximately equal to one tick, which essentially means that $M$ is so that $\alpha/2-\sigma/\sqrt{M}\geq 0$. Indeed, remind that the market spread is the smallest one achievable so that the market makers profit $S/2-\sigma/\sqrt{M}$ is non negative. Therefore, when decreasing the tick size, the market spread remains equal to one tick until the profit and loss of the market makers is equal to zero.

Using the fact that $\sigma$ is invariant when changing the tick value, together with the preceding assumptions and Equation \eqref{reg} where the spread is approximated by the tick value, we get
$$p_1\eta\alpha\sqrt{M}+p_2\alpha\sqrt{M}=p_{1,0}\eta_0\alpha_0\sqrt{M_0}+p_{2,0}\alpha_0\sqrt{M_0}.$$
Then, from the shape of the function $f$, we obtain the following formula:
\begin{equation*}
\eta\sim \Big(\frac{p_{1,0}\eta_0+p_{2,0}}{p_1}\Big)\Big(\frac{\alpha_0}{\alpha}\Big)^{1-\beta/2}-\frac{p_2}{p_1}.
\end{equation*}
Now, using different assumptions on $p_1$ and $p_2$, we obtain three versions of the prediction formula for the parameter $\eta$
after a change in the tick value. These three versions give the same order of magnitude for $\eta$, which is what matters in practice.
For the first version we assume $p_1=p_{1,0}$ and $p_2=p_{2,0}$. This gives the following result:
\paragraph{Effect of a change in the tick value on the microstructure, version 1:}
\begin{equation}\label{ti1}
\eta\sim \Big(\eta_0+\frac{p_{2,0}}{p_{1,0}}\Big)\Big(\frac{\alpha_0}{\alpha}\Big)^{1-\beta/2}-\frac{p_{2,0}}{p_{1,0}}.
\end{equation}
Again, recall that typical values for $\beta$ are $1$ and $1/2$.

If one wants to get even simpler formulas (which can be particularly important for regulators), which do not need any regression but still give the right order of magnitude, one may consider $p_1=p_{1,0}=1$ and $p_2=p_{2,0}=0.1$ or even $p_1=p_{1,0}=1$ and $p_2=p_{2,0}=0$. In these cases, we obtain the following simple results:
\paragraph{Effect of a change in the tick value on the microstructure, version 2:}
\begin{equation}\label{ti2}
\eta\sim (\eta_0+0.1)\Big(\frac{\alpha_0}{\alpha}\Big)^{1-\beta/2}-0.1.
\end{equation}
\paragraph{Effect of a change in the tick value on the microstructure, version 3:}
\begin{equation}\label{ti3}
\eta\sim \eta_0\Big(\frac{\alpha_0}{\alpha}\Big)^{1-\beta/2}.
\end{equation}

\begin{figure}
\centering
\includegraphics[width=\textwidth]{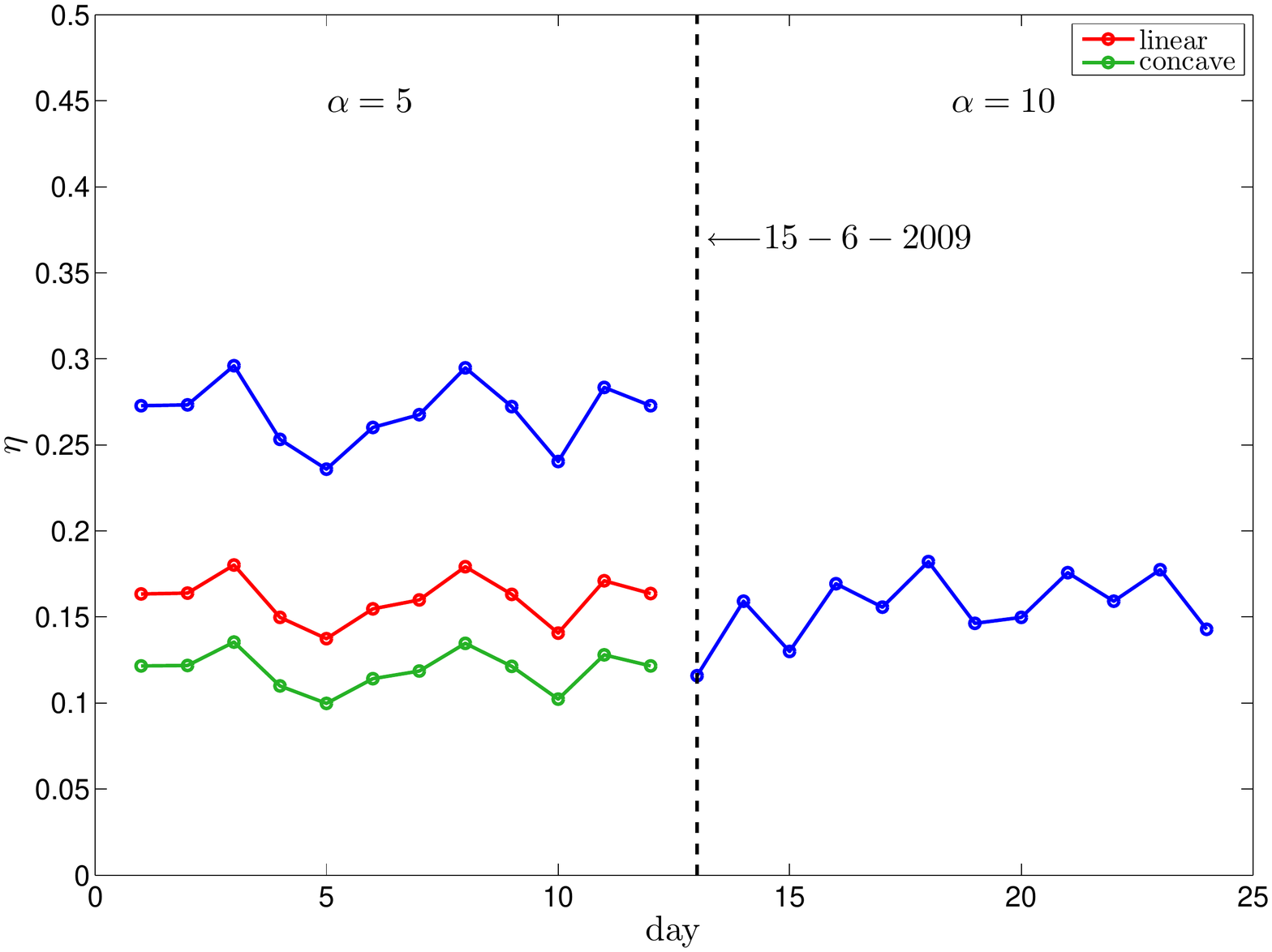}
\caption{Testing the prediction of $\eta$ on the Bobl futures. The blue lines show the daily measures of $\eta$. The red and green lines are the daily predictions associated to the future tick value.}\label{fig:etaBeforeAfter}
\end{figure}

Therefore, under reasonable assumptions, we are able to forecast the value of $\eta$ after a change in the tick value. In order to check these formulas on real data, we use the Bobl contract. The tick value of this asset has been multiplied by two on 2009, June 15. For 12 trading days before 2009, June 15, we give in Figure \ref{fig:etaBeforeAfter} the estimates of the value of $\eta$ after the change of the tick value given by Equation \eqref{ti1} (version 1 above) with $\beta=1$ and $\beta=1/2$.

The results are very satisfying, both assumptions on the latent liquidity leading to good estimates of the future value of $\eta$.

\subsection{Optimal tick value and optimal tick size}\label{optick}

Defining an optimal tick value is a very complicated issue, see for example \cite{ang}. Indeed, different types of market participants can have opposite views on what is a good tick value. We believe that our approach enables us to suggest a reasonable notion of {\it optimal tick value}. Of course the optimality notion we are about to define is arguable but still we think it is a first quantitative step towards solving the tick value question.

We consider that a tick value is optimal if:

\begin{itemize}
\item The (average) ex post cost of a limit order is equal to the (average) ex post cost of a market order, both of them equal to zero.
\item The spread is stable and close to one tick.
\end{itemize}

Such a situation can be seen as reasonable for both market makers and market takers. Indeed, it removes any implicit costs or gains due to the microstructure. Moreover, having a stable spread close to one tick prevents sparse order books which can drive liquidity away.

It is easy to see that getting an optimal tick value is equivalent to have $\eta=1/2$ together with a spread which is still equal to one tick. Thus, we refer to this last situation as the {\it optimal tick size} case. Note that the
optimal tick size is the same for any asset $(\eta=1/2)$, whereas the optimal tick value depends on the features of the asset.
Remark that in the optimal situation, the following properties follows for the microstructure:

\begin{itemize}
\item The last traded price can be seen as a sampled Brownian motion.
\item Consequently, the signature plot is flat.
\end{itemize}

Starting from a large tick asset, our approach enables us to reach the optimal tick size situation. Indeed, this is possible to obtain $\eta=1/2$ and a spread close to one tick by changing the tick value only assuming that $\eta$ increases continuously when the tick value decreases. Then, when modifying the tick value, the spread remains equal to one tick as long as $\alpha/2-\eta\alpha\geq 0$. Indeed, if $\alpha^*$ denotes the largest tick value such that $\eta=1/2$, then for all $\alpha>\alpha^*$, market makers make positive profits with a spread of one tick and consequently maintain this spread. Then, from Equations \eqref{ti1}, \eqref{ti2} and \eqref{ti3}, we obtain three versions of the formula for the optimal tick value leading to $\eta=1/2$:
\paragraph{Optimal tick value formula, version 1:}
\begin{equation}\label{optform}
\alpha\sim \alpha_0\Big(\frac{\eta_0p_{1,0}+p_{2,0}}{p_{1,0}/2+p_{2,0}}\Big)^{\frac{1}{1-\beta/2}}.
\end{equation}

\paragraph{Optimal tick value formula, version 2:}
\begin{equation}\label{optform2}
\alpha\sim \alpha_0\Big(\frac{\eta_0+0.1}{0.6}\Big)^{\frac{1}{1-\beta/2}}.
\end{equation}

\paragraph{Optimal tick value formula, version 3:}
\begin{equation}\label{optform3}
\alpha\sim\alpha_0(2\eta_0)^{\frac{1}{1-\beta/2}}.
\end{equation}

Of course we do not pretend that in practice, applying such rules will exactly lead to an optimal tick value (in our sense). However, we do believe that these simple formulas give the right order of magnitude for the relevant tick value of a given asset. To end this section, we give in Table \ref{tab:optvalues} the optimal tick values for our assets computed from Equation \eqref{optform} with  $\beta=1$ or $\beta=1/2$, and the average values of $\eta$ given in Table \ref{tab:intro:Stats1}. Note that these tick values are obtained from our 2009 database, and could be updated with more recent data.

\begin{table}[h!]
\begin{small}
\centering
\begin{tabular}{rccc}
\toprule
\multicolumn{1}{c}{Futures} & \multicolumn{1}{c}{Tick Value} &  \multicolumn{1}{c}{Optimal tick value} & \multicolumn{1}{c}{Optimal tick value} \\
&&  \multicolumn{1}{c}{$\beta=1$} & \multicolumn{1}{c}{$\beta=1/2$} \\
\midrule
\multicolumn{1}{l}{BUS5}   & 7.8125 \$& 2.7 \$ & 3.8 \$  \\
\multicolumn{1}{l}{DJ}  &  5.00 \$& 1.6 \$ & 2.3 \$\\
\multicolumn{1}{l}{EURO} & 12.50 \$& 3.1 \$ & 5.0 \$ \\
\multicolumn{1}{l}{SP}  &  12.50 \$& 0.3 \$ & 0.9 \$  \\\hline
\multicolumn{1}{l}{Bobl 1}  & \euros{5.00} & \euros{1.8} & \euros{2.6}  \\
\multicolumn{1}{l}{Bobl 2}   & \euros{10.00} & \euros{1.6} & \euros{2.8}  \\
\multicolumn{1}{l}{Bund}  &  \euros{10.00} & \euros{1.6} & \euros{2.9} \\
\multicolumn{1}{l}{DAX}  & \euros{12.50} & \euros{4.9} & \euros{6.7} \\\hline
\multicolumn{1}{l}{ESX}  &  \euros{10.00} & \euros{1.3} & \euros{2.6} \\
\multicolumn{1}{l}{Schatz}  & \euros{5.00} & \euros{0.8} & \euros{1.5} \\
\multicolumn{1}{l}{CL}  & 10.00 \$& 3.1 \$ & 4.6 \$ \\
\bottomrule
\end{tabular}%
\caption{Optimal tick values for the considered assets, in the linear case and the square root concave case.}\label{tab:optvalues}
\end{small}
\end{table}
Thus, according to our approach, tick values should be quite significantly reduced for the considered assets. It is particularly interesting to remark that the optimal tick values suggested for the Bobl are almost the same before and after the change in the tick value on 2009, June 15.

\subsection{Optimal tick value for small tick assets}

A crucial point in our approach is that when changing the tick value of a large tick asset, the spread remains equal to one tick as long as market makers make profit with such a spread. So the spread (in ticks) is invariant when the tick value is modified. For a small tick asset, when enlarging the tick value, both the spread and the number of trades adjust so that the spread and the volatility per trade have of the same order of magnitude. The way these two variables are jointly modified is intricate and this is why our method cannot, a priori, be used for small tick assets. However, let us stress the fact that it is still possible for the exchange to use a two step procedures in the case of a small tick asset:

\begin{itemize}
\item Step 1: Enlarge sufficiently the tick value so that the asset becomes a large tick asset.
\item Step 2: Use our methodology for large tick assets.
\end{itemize}

\section{Conclusion}

To conclude, we recall here the main messages of our paper, which applies to any large tick asset.

\begin{itemize}
\item {\it Measuring microstructure.} Most of the microstructure phenomena can be measured through a single parameter $\eta$, which is very easy to compute in practice. Therefore, to predict microstructure features after a change in the tick value, it suffices to predict the value of $\eta$ after such a change.
\item {\it Quantifying the tick size: the implicit spread.} In particular, $\eta$ quantifies the tick size of a large tick asset. Indeed, it can be shown both empirically and theoretically that the quantity $\eta\alpha$ plays the role of spread for large tick assets.
\item {\it Forecasting microstructure after a change in the tick value.} We obtain easy and straightforward formulas in order to forecast $\eta$, see Formulas \eqref{ti1} to \eqref{ti3}. These results are confirmed in practice thanks to spectacular results on the Bobl.
\item {\it Choosing a tick value leading to given microstructure effects.} Inversely, we can find the required $\alpha$ for any choice of $\eta$. In particular, it allows us to obtain simple explicit rules in order to reach what we call an optimal tick value.
\item {\it Optimal tick value.} In our approach, an optimal tick value is a tick value such that the ex post cost of a limit order is equal to the ex post cost of a market order, both of them equal to zero, and the spread is close to one tick.
\item {\it Optimal tick size.} Getting an optimal tick value corresponds to the case where the spread is close to one tick and the tick size is optimal, that is $\eta=1/2$. This situation can always be reached after a suitable change in the tick value of a large tick asset, see Formulas \eqref{optform} to \eqref{optform3}.
\end{itemize}

\section*{Acknowledgements}
We are grateful to R. Almgren, E. Bacry, J.P. Bouchaud, M. Hoffmann, J. Kockelkoren, C.A. Lehalle, J.F. Muzy, C.Y. Robert and S. Takvorian for helpful discussions.

\end{document}